\documentclass[11pt,a4paper,twoside,groupcitations]{article}
\usepackage[T1]{fontenc}
\usepackage[ansinew]{inputenc}
\usepackage[english]{babel}
\usepackage{amsfonts}
\usepackage{amsmath}
\usepackage{bm}
\usepackage{array}
\usepackage{amsthm}
\usepackage{amssymb}
\usepackage{graphicx}
\usepackage{subfigure}
\usepackage{braket}
\usepackage{eucal}
\usepackage{verbatim}
\usepackage[table]{xcolor}
\usepackage{caption}
\usepackage{cite}
\usepackage{textcomp}
\usepackage{hyperref}
\usepackage{multicol}
\usepackage{tikz}
\usetikzlibrary{positioning,arrows}
\usetikzlibrary{decorations.pathmorphing}
\usetikzlibrary{decorations.markings}
\usetikzlibrary{calc,decorations.markings}
\usetikzlibrary{arrows,shapes}
\usetikzlibrary{matrix,arrows}
\usepackage{pgfplots}
\usepackage{xparse}
\definecolor{jade}{HTML}{00A86B}
\newcommand{\be}{\begin{eqnarray}}
\newcommand{\ee}{\end{eqnarray}}

\newcommand{\expec}[1]{\mbox{$\langle\, #1\,\rangle$}}

\renewcommand{\d}{\mbox{${\rm d}$}} 
\newcommand{\lp}{\ell_{\rm p}}
\newcommand{\mpl}{m_{\rm p}}
\newcommand{\gn}{G_{\rm N}}

\newcommand{\Rh}{R_{\rm H}}

\numberwithin{equation}{section}
\title{\bf Quantum black hole ringdown}
\author{Konstantinos~Topaloglou$^{a}$\thanks{E-mail: kon.topaloglou@gmail.com},
Elena~Cuoco$^{ab}$\thanks{E-mail: elena.cuoco@unibo.it},
and
Roberto~Casadio$^{abc}$\thanks{E-mail: casadio@bo.infn.it}
\\
\\
$^a${\em Dipartimento di Fisica e Astronomia, Universit\`a di Bologna}
\\
{\em via Irnerio~46, 40126 Bologna, Italy}
\\
\\
$^b${\em I.N.F.N., Sezione di Bologna, I.S.~FLAG}
\\
{\em viale B.~Pichat~6/2, 40127 Bologna, Italy}
\\
\\
$^c${\em Alma Mater Research Center on Applied Mathematics (AM$^2$)}
\\
{\em Via Saragozza 8, 40123 Bologna, Italy}
}
\begin{document}
\maketitle
\begin{abstract}
The ringdown of a black hole following a merger is a potential candidate for revealing the signatures
of quantum gravity in the emerging gravitational waves.
In quantum theory, black holes are expected to have a discrete area and energy spectrum,
which conflicts with the classical notion of an horizon that absorbs all infalling perturbations.
We propose that the quantum black hole dissipates the energy it cannot absorb
by emitting ``soft'' gravitons that carry away the energy difference between the energy of the
infalling perturbation and the quantum transition energy.
We find that the ringdown spectrum is consequently enriched with low-frequency components,
and that there exists a weak low-frequency flux that persists for a timescale much longer
than the ringdown itself.
\end{abstract}
\section{Introduction}
\setcounter{equation}{0}
\label{S:intro}
In anticipation of a successful theory of quantum gravity, there has been much interest in predicting
whether and how such a theory could appreciably influence systems that can be observed.
One of such systems is given by astrophysical black holes~\cite{Fabian:2019sxb} in binary configurations
that generate strong gravitational wave signals during the merger~\cite{Schmidt:2020ekt,LIGOScientific:2016aoc}.
\par
In particular, black holes as quantum systems (qBHs) have been argued to possess features that are universal
in quantum physics.
Long ago, Bekenstein~\cite{Bekenstein:1973ur} interpreted the entropy associated with the horizon surface area as counting
the degeneracy of black hole quantum states, thus obtaining a quantisation law for the horizon area
$\mathcal A_{\rm H}=4\,\pi\,\Rh^2=16\,\pi\,\gn^2\,M^2$ for a Schwarzschild black hole~\cite{Schwarzschild:1916uq}
of mass $M$.
This conjecture results in the Bekenstein-Mukhanov spectrum~\cite{Bekenstein:1995ju}~\footnote{We
shall use units with $c=1$ and often write the Planck constant $\hbar=\lp\,\mpl$ and the Newton constant $\gn=\lp/\mpl$,
where $\lp$ and $\mpl$ are the Planck length and mass, respectively.}
\be
\left( \frac{M}{\mpl}\right)^2
=
\frac{\bar\alpha}{16\, \pi}\,n
\equiv
\frac{\ln \alpha}{4\,\pi}\,n
\ ,
\label{eq:Bek_muk_quantisation}
\ee
where $n>0$ is an integer that labels the energy level and $\bar\alpha=4\,\ln\alpha>0$ is a parameter that depends
on the details of the model.
For macroscopic black holes $M\gg\mpl$, and the difference between adjacent energy levels
$\Delta M / M\equiv \hbar\,\varpi/M \ll 1$.
In this regime, we then obtain that qBH energy levels differ by the Bekenstein-Mukhanov
quantum
\be
\varpi_{\rm BM}
=
\frac{\ln\alpha}{8\, \pi\,\lp}
\left(\frac{\mpl}{M}\right)
\ .
\label{eq:Bek_muk_quantum}
\ee
A discrete energy spectrum is generically expected for quantum bound states, and we will therefore assume
that a qBH is characterised by the spectrum~\eqref{eq:Bek_muk_quantisation} in the rest of this work.
Moreover, we will consider a range of values for $1<\alpha<10^3$, but we recall that 
$\alpha$ is expected to be an integer with $\alpha=2$ being the preferred value according to the arguments
given in Ref.~\cite{Bekenstein:1995ju}.
\par
This quantum picture is at odds with the classical notion of a black hole as an object
which swallows everything that falls through its event horizon, because bodies carrying an arbitrary
amount of energy $E$ could be absorbed by a qBH only if they induce excitations corresponding to 
transitions permitted by Eq.~\eqref{eq:Bek_muk_quantisation} (say from $n$ to $n'>n$, both being integer).
The easiest solution to this problem is to assume that the excess energy $\Delta E\simeq E-(n'-n)\,\hbar\,\varpi_{\rm BM}$
is emitted by the black hole and it thus becomes interesting to speculate how a qBH loses the excess
energy $\Delta E$ while transitioning from one energy level to another.
\par
This question can hardly be addressed during the merging of two black holes, when non-perturbative
effects dominate and the problem can only be studied numerically, but it becomes amenable to a perturbative
analysis during the subsequent ringdown phase, while the system settles down to the final black hole
configuration~\cite{Schmidt:2020ekt}.
The ringdown stage is characterised by quasinormal mode (QNM) oscillations, that is time-decaying solutions
of the linearised field equations in the (final) black hole exterior region with boundary conditions
such that there exists only an infalling component at the event horizon and an outgoing component
at radial infinity.
QNMs turn out to be characterised by a discrete set of complex-valued frequencies (different for each spin $s$;
see Ref.~\cite{Berti:2009kk} for a detailed review).
\par
To account for potential non-classical properties of the horizon, Ref.~\cite{Cardoso:2019apo} proposed to modify
the near-horizon behaviour of the perturbation modes by introducing a frequency-dependent reflective coefficient
for the horizon.
The energy carried by modes whose frequency does not correspond to the permitted energy gaps
(multiple of $\varpi_{\rm BM}$) is therefore reflected outwards and results in {\em echoes\/} of the primary
QNM signals~\cite{Cardoso:2001bb} (see also Refs.~\cite{Chakraborty:2022zlq,Coates:2021dlg,Maggio:2020jml,Maggiore:2007nq}).
In such an approach qBHs are not perfect absorbers and infalling particles will have a chance
of getting reflected outwards as they approach the black hole horizon.
\par
In this work we propose an alternative mechanism in which infalling perturbations with arbitrary energy $E$
will be absorbed by the qBH unconditionally, but the excess energy $\Delta E$ is emitted by the qBH
in the form of gravitational waves. 
This process differs from {\em echoes\/} in two ways:
first, the dissipation is performed via gravitons no matter what the type of infalling matter;
second, the frequency of the emerging gravitons will correspond to the energy increment $\Delta E$
(divided by $\hbar$).
This, in particular, implies that we do not modify the boundary condition at the horizon when
dealing with field perturbations, and the QNM frequencies remain unchanged.
Rather, the ringdown picture is enriched by the presence of an additional outgoing flux of gravitons.
\par
Our proposal is motivated by the behaviour of analogous quantum systems.
We can consider, for example, the capture of a free electron by the proton
leading to the formation of the hydrogen atom.
Within the approximation that the proton is much heavier and therefore essentially stationary,
we can view the electron as falling within the Coulomb potential of the proton.
The positive electric field accelerates the electron towards the proton and causes it to emit kinetic energy
in the form of photons.
This loss of energy reduces the electron energy from positive (free motion) to negative (bound motion)
and terminates precisely when the electron reaches the ground state energy level.
The excess energy here results in photons, the quanta of the electromagnetic field.
We view the capture of a (relatively small) object by a (relatively large) qBH as a similar process,
with the gravitational field (gravitons) naturally replacing the Coulomb field (photons). 
\par
In what follows, our goal is to illustrate how this process could take place during the ringdown of a
Schwarzschild black hole~\cite{Schwarzschild:1916uq}.
Although astrophysical black holes are expected to carry angular momentum~\cite{Fabian:2019sxb,Kerr:1963ud}, 
the spherical geometry is still sufficient to test our proposal and estimate some features of the gravitational wave
spectrum that should emerge.
By considering excited QNMs with some profile (which depends on the initial disturbance),
and by computing the relative probability of the transitions that they can induce, we will find
the frequencies and magnitudes of the emission lines of the dissipative gravitons.
As a working assumption, we will rely on the qBH model introduced in Ref.~\cite{Casadio:2021eio},
where a spectrum of the type in Eq.~\eqref{eq:Bek_muk_quantisation} is obtained by describing
the geometry with the coherent state of a scalar field whose expectation value reproduces
the Schwarzschild solution (with some quantum modifications resulting from regularisation).
We will finally evaluate the scattering amplitude for the capture process with in- and out-states containing
the qBH as a coherent state, as well as additional gravitons representing the infalling QNMs and the
dissipative gravitons.
\par
In Section~\ref{S:coBH}, we briefly review the qBH model from Ref.~\cite{Casadio:2021eio}
and obtain the relative transition probabilities by scattering processes of perturbations of the
gravitational field.
In Section~\ref{S:spectrum}, we apply our probability estimates to a set of QNMs following a given
excitation profile, and, by accounting for the greybody factors of the spacetime experienced by
the outgoing modes, obtain the enriched ringdown spectrum as seen by a distant observer.
In particular we observe that the low-frequency fluxes we anticipate have such low transmission
that they are effectively trapped in the near-horizon spacetime for a long time after the classical
ringdown and continue to leak at a small rate.
In Section~\ref{S:conc}, we comment on our results.
\section{Coherent qBH model}
\label{S:coBH}
The Schwarzschild line element for a black hole of mass $M$ is given by~\cite{Schwarzschild:1916uq}
\be
\d s^2
=
-\left(1+2\,\phi\right)\d t^2
+
\frac{\d r^2}{1+2\,\phi}
+
r^2\,\d\Omega^2
\ ,
\label{schw}
\ee
with $2\,\phi=-\Rh/r=-{2\,\gn\,M}/{r}$.
Following Ref.~\cite{Casadio:2021eio}, we can try to recover this metric from the quantisation of
$\phi$ as a massless scalar field in Minkowski spacetime and a quantum state $\ket M$ such that
\be
\expec{\hat \phi}
\equiv
\bra M \hat \phi \ket M
\simeq
\phi(r)
\ .
\label{eq:phi_expectation_val}
\ee
The metric~\eqref{schw} can indeed be reproduced by (normalised) coherent states
of $\hat\phi$ but only approximately, due to the mandatory normalisability of $\ket M$.
\par
The field $\phi$ must satisfy the Klein-Gordon equation
\be
\Box\phi
=
0
\ ,
\ee
where $\Box$ is the d'Alembert operator in flat spacetime.
In spherical coordinates, one can write the general solution as~\footnote{We denote the complex conjugate
of $f$ with $\bar f$.} 
\be
\hat\phi
=
\int \frac{k^2\,\d k}{2\, \pi^2} 
\left(\hat \alpha_k\, u_k + \hat \alpha_k ^\dagger\, \bar u_k\right)
\ ,
\ee
where the normal modes are given by
\be
u_k
\propto
e^{-i\, k\, t}\,\frac{\sin(k\,r)}{k\,r} 
\ ,
\label{u_k}
\ee
with the normalisation fixed by the Klein-Gordon scalar product
\be
(u_k, u_p)
=
4\,\pi\,i\int_0^\infty
r^2\,\d r
\left[
\bar u_k\left(\partial_t u_p\right)
-
\left(\partial_t \bar u_k\right)u_p
\right]
=
\frac{4\, \pi^2}{k^2}\, \delta(k-p)
\label{eq:norm_modes}
\ .
\ee
The field $\hat\phi$ and its conjugate momentum $\hat\Pi$ satisfy equal time commutation relations, 
\be
\left[ \hat \phi(t,r), \hat \Pi(t,s)\right]
=
\frac{i\, \hbar}{4\, \pi\, r^2}\,\delta(r-s)
\ ,
\ee
which yield the usual commutators
\be
\left[ \hat \alpha_k, \hat \alpha^\dagger_p\right]
=
\frac{2\, \pi^2}{k^2}\,\delta(k-p)
\ ,
\ee
for the ladder operators $\hat \alpha_k$ and $\hat \alpha^\dagger_k$. 
The quantum vacuum is then defined by $\hat a_k\ket 0=0$ for all $k>0$ and corresponds to
the vacuum Minkowski geometry~\cite{Casadio:2021eio}.
\subsection{Coherent state construction}
The state $\ket {M}$ in Eq.~\eqref{eq:phi_expectation_val} is taken from Ref.~\cite{Casadio:2021eio}
to be a coherent state created from the vacuum $\ket 0$ by the action of a unitary shift operator
$\hat D_g$, that is
\be
\ket {M}
=
\hat D_g \ket 0
\equiv
\exp{  \int_{0} ^\infty  \frac{k^2\,\d k}{2\, \pi^2}
\left(-\frac{|g(k)|^2}{2}+{g}_k\, \hat \alpha_k^\dagger \right)}
\ket 0
\ .
\label{M_coherent_state}
\ee
The complex function $g(k) = g_k\, \exp( i\, \gamma_k)$ of the wavenumber $k$ determines
the coherent state and is fixed by the requirement~\eqref{eq:phi_expectation_val},
which yields
\be
g_{\rm nr}(k)
=
-\frac{4\, \pi\,M}{\sqrt {2\, k^3}\,\mpl}\,e^{i\,k\,t}
\ .
\label{g_function_Schw}
\ee
The subscript ``nr'' (non-regularised) here is a reminder that the functions~\eqref{g_function_Schw}
must be regularised to define a proper coherent state.
In fact, the total occupation number turns out to be given by
\be
\expec{\hat N}_{\rm nr}
=
4
\left(\frac{M}{\mpl}\right)^2 
\int_0^\infty \frac{\d k}{k}
\ ,
\label{divN}
\ee
which diverges both in the infrared (IR) for $k\to 0$ and in the ultraviolet (UV) for $k\to \infty$.
Instead of modifying the functions~\eqref{g_function_Schw} by relying on a specific model,
we can introduce regularising cutoffs such that $0<k_{\rm IR}<k<k_{\rm UV}<\infty$, which yields a finite
total occupation number $\expec{\hat N}$
and
\be
\left(\frac{M}{\mpl}\right)^2
=
\frac{\expec{\hat N}}{4\,\ln\!\left({k_{\rm UV}}/{k_{\rm IR}}\right)}
\ .
\label{coherent_mass_reg}
\ee
In the limit $k_{\rm IR} \to 0$, one finds a modified classical solution of the form~\eqref{schw}
with the classical $\phi$ replaced by~\cite{Casadio:2021eio,Urmanov:2024qai}
\be
\phi_{\rm reg}
=
\frac{2}{\pi}\,\phi(r)\,\operatorname{Si}\!\left(\frac{r}{R_{\rm UV}}\right)
\ ,
\ee
where $R_{\rm UV} = 1/k_{\rm UV}$ and
\be
\operatorname{Si}(x) = \int_0 ^x \frac{\sin t}{t} dt
\ee
is the sine integral function.
\par
We note that Eq.~\eqref{coherent_mass_reg} is of the form of the Bekenstein-Mukhanov quantisation
law~\eqref{eq:Bek_muk_quantisation}, save for the fact that $\expec{\hat N}$ is not necessarily integer.
We can nonetheless assume that $\expec{\hat N}$ changes by integer increments any time matter
falls into the black hole.
The multiplicative factor can also be matched between the two descriptions by setting
\be
\frac{k_{\rm UV}}{k_{\rm IR}}
=
e^{\frac{\pi}{\ln\alpha}}
\ .
\label{reg_ratio_to_Bek_Muk}
\ee
In the following, we will employ this description of qBH for practical calculations given that it entails the general
feature $M^2 \sim \expec{\hat N}$ for $g(k) \sim M$.
\subsection{Scattering with an infalling perturbation}
\label{SS:scat}
Our goal is now to compute the probability that an infalling perturbation of energy $E$
can excite the qBH of mass $M$ to a higher state of mass $M+\delta M$ in the Bekenstein-Mukhanov
spectrum~\eqref{eq:Bek_muk_quantisation} and dissipate the difference $\Delta E=E-\delta M$ by emitting
gravitons.
Since we model the qBH as a coherent state of a quantum field, it is natural to describe
the excitation as a scattering process within the standard framework of quantum field
theory.
We must however take particular care in specifying the past and future asymptotic
states $\ket{\rm in}$ and $\ket{\rm out}$ for the infalling perturbation and outgoing gravitons
that take part in the scattering.
\par
The simplest idealisation would be to define $\ket{\rm in}$ and $\ket{\rm out}$ when the qBH and
the localised perturbation are far enough that the spacetime is approximately flat at the location
of the perturbation.
Although we can always consider a lone particle approaching from infinity in the state $\ket{\rm in}$,
the state $\ket{\rm out}$ will necessarily contain outgoing gravitons that backscatter in the
background potential.
The future asymptotic state therefore cannot be made up of particles separated enough
from the qBH that they can be considered asymptotically free.
\begin{figure}[t]
\centering
\includegraphics[width=0.7\linewidth]{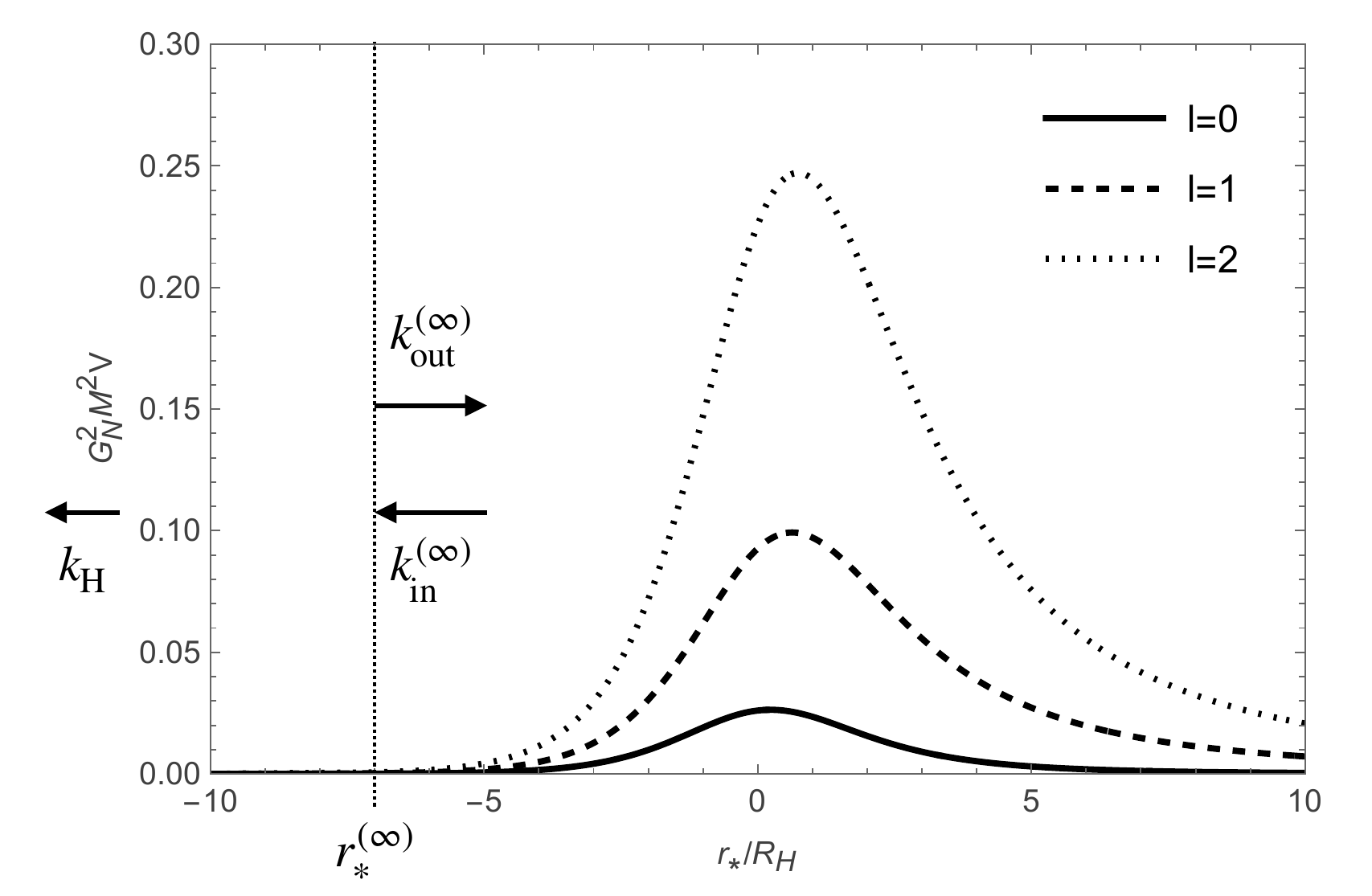}
\caption{Potential for massless scalar perturbations in the outer Schwarzschild spacetime
in tortoise coordinates $r_*=r+\Rh\,\ln(r/\Rh-1)$ (perturbations of spin $s=2$ experience the similar Regge-Wheeler and Zerilli
potentials). 
Asymptotic particle states are defined at $r_*^{(\infty)}=r_*(\Rh+L)$ with $\lp\ll L\ll \Rh$
with purely ingoing boundary condition at $r_*(\Rh)=-\infty$,
so that the scattering occurs in the region $-\infty<r_*<r_*^{(\infty)}$.}
\label{fig:potential}
\end{figure}
\par
A more convenient approach is to consider that the asymptotic past and future refer to the outer 
boundary of a near-horizon region, $\Rh<r<\Rh+L$, with $\lp \ll L \ll R_{\rm H}=2\,\gn\,M$,
where the propagation of modes is not significantly affected by the 
potential due to the curved background (see Fig.~\ref{fig:potential} for the case of scalar perturbations).
This assumption is motivated by the observation that one does not anticipate effects of quantum gravity
at a large distance from the horizon, but also that any such effect that could take place beyond the horizon
should remain undetected by external observers.~\footnote{For example, in Ref.~\cite{Calmet:2021stu}
quantum corrections to the geometry were shown to be highly suppressed by factors of $(\lp/\Rh)^2$.
Possibly larger modifications to the QNM spectra are obtained in Refs.~\cite{Bambagiotti:2025qxj,Antonelli:2025yol},
but the main predictions of the present work would remain qualitatively unaffected.}
For a reasonably large black hole (of at least a few solar masses) there is plenty of freedom in defining
the length scale $L$.
Asymptotic states for the (ingoing and outgoing) particles are therefore defined at the times
$t_{-\infty} < t_{+\infty}$ when the particles are at $r^{(\infty)}=\Rh+L$. 
We remark that this approach requires that the background for the particle scattering be given
by the geometry of the qBH of initial mass $M$, while infalling particles carrying energy $E$
have not yet been absorbed by the qBH at the time $t_\infty$ when the excess energy is emitted. 
\par
The past and future asymptotic states of the qBH are going to be given by the coherent states
$\ket M$ and $\ket{M+\delta M}$, respectively, where the mass increment $\delta M = n \,\hbar\, \varpi_{\rm BM}$.
We therefore write the total transition amplitude as~\cite{Skaar2023} 
\be
\mathcal M
=
\bra{{\rm out} ; M + \delta M}S[\hat\phi] \ket{{\rm in} ; M}
\ ,
\label{S_matrix_element}
\ee 
where $S$ is the S-matrix which acts between the asymptotic state $\ket {\rm in}$
containing the infalling perturbation of energy $E$ and the asymptotic state $\ket {\rm out}$
containing the dissipative gravitons defined at the location detailed above and carrying the excess
energy $\Delta E$.
Note that we will later consider the propagation of the gravitons in the asymptotic state $\ket{\rm out}$
through the potential barrier in order to reconstruct the actual signal detected far from the qBH.
\par
We can now use the properties derived in Ref.~\cite{Ilderton:2017xbj} to extract the coherent states
from the amplitude.
The operators $\hat D_g$ acting on the asymptotic qBH states lead to an exponential factor
and shifts the fields $\hat \phi$ inside the S-matrix by a background configuration
determined by the functions $g_M(k)$ and $g_{M+\delta M}(k)$ corresponding
to the initial and final coherent states respectively, that is
\be
\mathcal M 
&\!\!=\!\!&
e^{\strut\displaystyle -\frac{1}{2}
\int \frac{k^2\,\d k}{2\, \pi^2}
\left[ |g_M(k)|^2 +|g_{M+\delta M}(k)|^2 - 2\,\bar g_{M+\delta M}(k)\,g_M (k) \right]}
\nonumber
\\
&&
\times
\bra{\rm out}
e^{\strut\displaystyle\int \frac{k^2\,\d k}{2\, \pi^2}
\left[g_M(k)\, \hat \alpha^\dagger_k \right]}
S[\hat \alpha + g_M, \hat \alpha^\dagger + \bar g_{M+\delta M}]
e^{\strut\displaystyle\int \frac{k^2\,\d k}{2\, \pi^2}
\left[\bar g_{M+\delta M}(k)\, \hat \alpha_k \right]}
\ket{\rm in}
\ ,
\qquad
\ee
where the fields in the S-matrix have been shifted as
$\hat \alpha_k \to \hat \alpha_k + g_M(k)\, \hat{\mathbb I}$
and $\hat \alpha^\dagger_k \to \hat \alpha^\dagger_k + \bar g_{M+\delta M}(k)\,\hat{\mathbb I}$.
Taking the square modulus of $\mathcal M$, we find the probability density
\be
\mathcal P(\delta M)
=
e^{\strut\displaystyle-4\,\ln\!\left( \frac{k_{\rm UV}}{k_{\rm IR}}\right) \frac{\delta M^2}{\mpl^2}}\,
\left| \mathcal M'\right|^2
\ .
\ee
Invoking the matching~\eqref{reg_ratio_to_Bek_Muk}, and recalling the mass increment
$\delta M = n\, \hbar\, \varpi_{\rm BM}$, we obtain
\be
\mathcal P(n)
=
e^{\strut\displaystyle-\frac{\mpl^2\,\ln\alpha}{16\,\pi\,M^2}\,n^2}\,
\left| \mathcal M'\right|^2
\ .
\label{P(n)}
\ee
The remaining matrix element,
\be
\mathcal{M}'
=
\bra{\rm out}
e^{\strut\displaystyle\int \frac{k^2\,\d k}{2\, \pi^2}
\left[g_M(k)\, \hat \alpha^\dagger_k \right]}
S[\hat\phi + \phi_{\rm bkg}]\,
e^{\strut\displaystyle\int \frac{k^2\,\d k}{2\, \pi^2}
\left[\bar g_{M+\delta M}(k)\, \hat \alpha_k \right]}
\ket{\rm in}
\label{amplitude_element_withexpon}
\ee
is evaluated on the composite background 
\be
\phi_{\rm bkg}
=
\int \frac{k^2\,\d k}{2\, \pi^2} 
\left[g_M(k)\,u_k + \bar g_{M+\delta M} (k)\, \bar u_k\right]
=
\phi_M
+
\int \frac{k^2\,\d k}{2\, \pi^2}
\left[\bar g_{M+\delta M}(k) - \bar g_M(k)\right]
\bar u_k 
\ .
\ee
Note that this expression is real due to the form of the normal modes~\eqref{u_k}
and corresponds to an intermediate configuration 
\be
\phi_{\rm bkg}
=
\frac{1}{2}
\left(\phi_{\rm bkg} + \bar \phi_{\rm bkg}\right)
=
\frac{1}{2}
\left(\phi_{M} + \phi_{M+\delta M}\right)
\ .
\ee
\par
To complete the calculation, we model the infalling perturbation with a single QNM of frequency
$\omega_{\rm QNM}$ in the state $\ket {\rm in}$ and assume that the state $\ket{\rm out}$
is a single outgoing graviton with the appropriate frequency corresponding to the excess energy
$\omega_{\rm QNM} - \delta M/\hbar$.
These states are spherical waves $\ket{\psi}=\hat A_p^\dagger\ket{0}$ with angular momentum
$l \geq 2$ created from the (locally flat) vacuum $\ket 0$ by creation operators
\be
\hat A_p^\dagger
=
\sum_{l\ge 2}
\int \frac{k^2\,\d k}{2\,\pi^2}\,
\psi_l(p,k)\, \hat \alpha_{l,k}^\dagger
\ ,
\ee
where $\psi_l(p,k)$ are suitable profile functions.
The ladder operators $\hat \alpha_{l\ge 2}$ and $\hat \alpha_{l\ge 2}^\dagger$
commute with the $\hat\alpha$ and $\hat\alpha^\dagger$ that annihilate and create monopole modes.
The exponential insertions of monopole ladder operators in the amplitude~\eqref{amplitude_element_withexpon}
therefore reduce to the identity when acting on $\ket{\rm in}=\ket{\psi}\ket{0_{l=0}}$,
\be
e^{\strut\displaystyle\int \frac{k^2\,\d k}{2\, \pi^2}
\left[\bar g_{M+\delta M}(k)\, \hat \alpha_k \right]}
\ket{\rm in}
=
\ket{\psi}\,
e^{\strut\displaystyle\int \frac{k^2\,\d k}{2\, \pi^2}
\left[\bar g_{M+\delta M}(k)\, \hat \alpha_k \right]}
\ket{0_{l=0}}
=
\ket{\psi}
\ket{0_{l=0}}
\ ,
\ee
and similarly for $\bra{\rm out}$.
We therefore only need to compute
\be
\mathcal M'
=
\bra{\rm out}
\hat S[\hat \phi + \phi_{\rm bkg}]
\ket{\rm in}
\ .
\label{amplitude_element}
\ee
The quantity of interest is the dependence of the amplitude on the number of excitations
$n$ in $\delta M = n\, \hbar\, \varpi_{\rm BM}$, which itself only appears as part of the shifted
background $\phi_{\rm bkg}$. 
\par
The state $\ket{\rm out}$ could just contain a single outgoing graviton or a plethora
of gravitons trapped by the background potential, which would result in the interaction vertices
relevant for the process.
In any case we expect that the result will separate as
\be
\mathcal M'
=
(\text{vertex})\times (\text{4-momentum conservation})
\ee
and that it should be possible to Taylor expand the vertex factor in the small parameter
$\delta M / M$.
We make no explicit choice here, but observe that a vertex value of leading $\mathcal O(1)$ 
will lead to
\be
\mathcal P_1(n)
\sim
e^{\strut\displaystyle-\frac{\mpl^2\,\ln\alpha}{16\, \pi\,M^2}\,n^2} 
\label{eq:prob_profile_1}
\ ,
\ee
whereas one of leading $\mathcal O(\delta M/M)=\mathcal O(n)$ yields
\be
\mathcal P_2(n)
\sim
n^2\,e^{\strut\displaystyle-\frac{\mpl^2\,\ln\alpha}{16\, \pi\,M^2}\,n^2}\, 
\ ,
\label{eq:prob_profile_2}
\ee
with the normalisations that will be fixed by summing over all possible excitations and setting
the total probability to 1. 
We note that in practice the quantity in the exponent is very small for any reasonable
$n\sim \delta M\ll M$ and choice of $\alpha$, so that $\mathcal P_1 \sim 1$ and $\mathcal P_2 \sim n^2$.
\section{The dissipation spectrum}
\label{S:spectrum}
We now use the probability densities~\eqref{eq:prob_profile_1} and~\eqref{eq:prob_profile_2}
in the context of the BH ringdown, where we assume that a variety of QNMs have been excited
with some arbitrary profile.
For perturbations with $l=2$ we have a set of complex QNM frequencies $\{\sigma_N=\omega_N - i \,\gamma_N\}$
with increasing imaginary part $\gamma_N$ for increasing integer $N$~\cite{Berti:2009kk}.
The excitation profile following a realistic merger event or other types of disturbance is not known,
but we will assume that the excitation contains primarily the fundamental $N=0$ mode and to a lesser
degree the rest of the modes, suppressed for increasing $N$.
The ringdown signal of the field $\phi(t,r)$ observed at a location $r=r_{\rm obs}$ is thus modelled as
\be
\psi_{\rm obs}(t, r=r_{\rm obs})
=
A\,\cos(\omega_0\,t)\,e^{-\gamma_0\,t}
+
\frac{A}{2}\,\cos( \omega_1\,t)\,e^{-\gamma_1\,t}
+
\frac{A}{10}\,\sum_{N \ge 2}
e^{-N}\,\cos(\omega_N\, t)\, e^{-\gamma_N\, t}
\ ,
\label{excitation_profile}
\ee
with a global amplitude $A$.
\par
The near-horizon spectrum is obtained as follows.
Each excited QNM is assumed to contain infalling quanta of energy $E_N = \hbar \,\omega_N$
which scatter with the qBH of mass $M$ and induce excitations to any mass levels $M+\delta M$
characterised by $n$ such that $\Delta E=E_N -\delta M=E_N - n\, \hbar\, \varpi_{\rm BM} \ge 0$.
For each of the $\max{\{n;E_N\}}$ possibilities we compute $\mathcal P(n)$ and the
relative probability
\be
P(n;E_N)
=
\frac{\mathcal P(n)}{\sum\limits_{n' \le \max{\{n;E_N\}}} \mathcal P(n')}
\ .
\label{P(nE)}
\ee
As a result we obtain new spectral lines corresponding to the frequencies $\Delta E/\hbar=E_N/\hbar - n\,\varpi_{\rm BM}$,
with decay rate equal to the original $\gamma_N$, and amplitude determined by weighting the original QNM
power by $P(n;E_N)$.
\subsection{Near-horizon greybody factors}
The spectral lines that we obtained above represent gravitons that are emitted near the horizon
and therefore need to travel through the exterior spacetime in order to reach a distant observer
who will measure their flux as a gravitational wave.
This flux will be reduced because of the backscattering caused by the
Regge-Wheeler-Zerilli effective potentials~\cite{Chandrasekhar:1985kt} which peak near the light ring
(at $r\simeq 3\,\Rh/2\equiv R_{\rm lr}$).
The amplitude of the dissipative spectral lines must therefore be weighted by the transmission
factor $T=T(\omega)$ that we will just study for the even-parity modes satisfying the Zerilli
equation~\cite{Zerilli:1970se}~\footnote{The isospectrality of QNM of the Schwarzschild
spacetime~\cite{Chandrasekhar:1975nkd,Chandrasekhar:1975zza,Chandrasekhar:1985kt}
allows for a similar analysis of odd-parity perturbations.}
\be
\left[\partial^2_{x} + \omega^2- V_{\rm Z}(x)\right]\psi = 0 
\ ,
\ee
where $x = r_*/\gn\,M$, with $r_*=r+\Rh\,\ln(r/\Rh-1)$.
The dimensionless Zerilli effective potential is given by $V_{\rm Z}(x)=(\gn\,M)^2\,V_{\rm Z}(r)$, 
where
\be
V_{\rm Z}(r)
=
\left(1-\frac{2\,\gn\,M}{r}\right)
\frac{2\,\xi^2\,(\xi+1)\,r^3+6\,\xi^2\,\gn\,M\,r^2+18\,\xi\,\gn^2\,M^2\,r+18\,\gn^3\,M^3}
{r^3\left(\xi\,r+3\,\gn\,M\right)^2}
\ ,
\label{Zerilli_Veff}
\ee
with $2\,\xi=(l-1)\,(l+2)$.
\par
We can estimate the transmission factor by considering a stationary beam across the potential
Zerilli barrier.
The potential $V_{\rm Z}$ vanishes for $x \to \pm \infty$, so that the solutions in these regions
can be approximated by free waves.
We consider a stationary beam incoming from the left of the barrier from a starting point $x_{0}\ll-1$,
\be
\psi(x\sim x_0)
=
e^{-i\,\omega\, x}
+
\mathcal R_\omega\, e^{i\,\omega x}
\ee
and an outgoing mode to the right of the barrier at $x_1\gg 1$,
\be
\psi(x\sim x_1)
=
\mathcal T_\omega\, e^{-i\omega x} 
\ .
\ee
We solve for $\psi=\psi(x)$ across the potential barrier numerically up to $x_1 \gg 1$,
and fit into right- and left-moving modes to find the transmission factor $T(\omega) = |\mathcal T_\omega|^2$.
\par
We can also study the propagation of a wavepacket through the Zerilli barrier.
We consider an initial configuration of the field representing a localised 
right-moving wavepacket centred around the target frequency $\omega$, as
\be
\phi_\omega(t=0,x)
=
\frac{e^{-\frac{(x-x_{0})^2}{2\,\sigma^2}}}{\sqrt{2\, \pi}\,\sigma}\,
\cos(\omega\, (x-x_{0}))
\label{wavepacket_init_config}
\ee
and
\be
\dot \phi_\omega (t=0,x)
=
\frac{e^{-\frac{(x-x_{0})^2}{2\,\sigma^2}}}{\sqrt{2\, \pi}\,\sigma}\,\omega
\left[
\sin( \omega\, (x-x_{0})) + \frac{x-x_{0}}{\omega\,\sigma^2}\, \cos( \omega \,(x-x_{0}))
\right]
\ ,
\label{wavepacket_init_der}
\ee
where $\sigma$ is the width of the wavepacket and $x_{0}$ the initial central location.
The time derivative is chosen to satisfy the condition $\dot \phi_\omega (0,x)=-\partial_{x} \phi_\omega(0,x)$
making the wavepacket right-moving in the free region.
Numerically evolving this configuration in time on a grid by running fourth-order Runge-Kutta algorithm~\cite{Hairer:1993},
we locate the transmitted wavepacket after it has crossed the barrier and obtain the transmission factor
by measuring its amplitude.
The latter is computed by recording the signal $\phi_\omega(t,x_{\rm obs})$ that is observed at a fixed
location $x_{\rm obs}$, taking its Hilbert transform to find the envelope and locating the peak of the envelope.
\begin{figure}[t]
\centering
\includegraphics[width=0.7\linewidth]{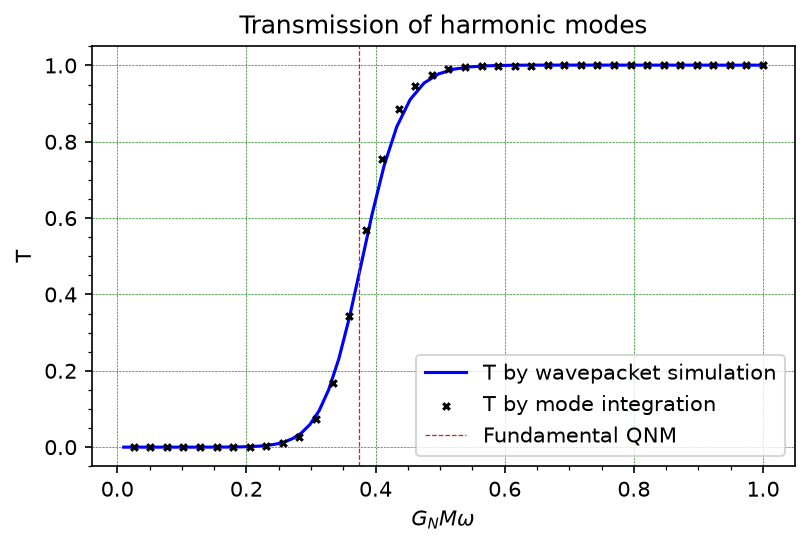}
\caption{Transmission factor for the Zerilli potential obtained by integrating a stationary mode across
the barrier (black markers), and by simulating $\sigma = 40$ wavepackets scattering on the barrier
(blue line).
The transmission factor $T$ in both cases is defined as the squared modulus of the ratio of the transmitted wave amplitude
to the incident wave amplitude.}
\label{fig:transmission_factor}
\end{figure}
\par
\begin{figure}[t]
\centering
\includegraphics[width=0.8\linewidth]{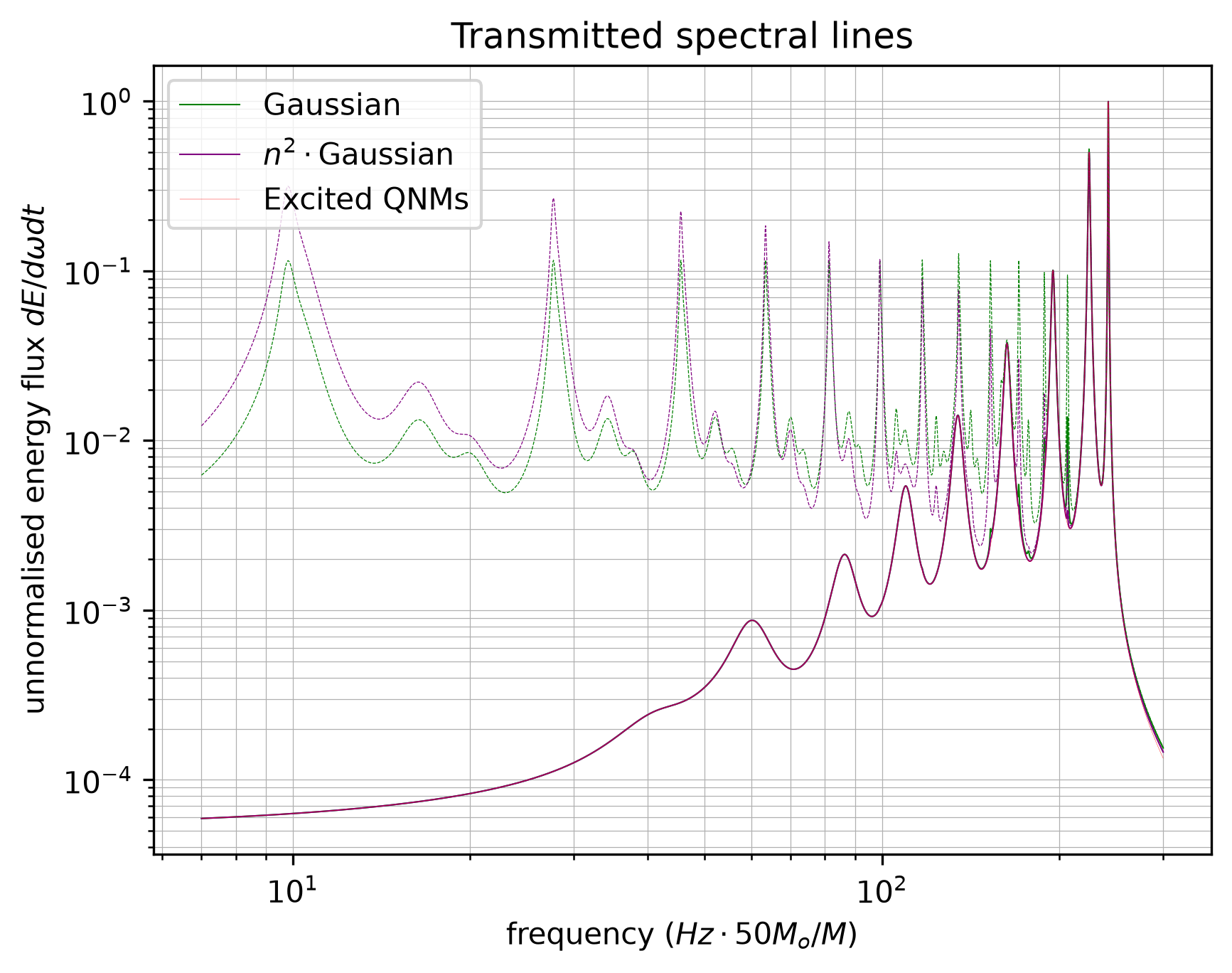}
\caption{Emitted spectrum for the probability profiles~\eqref{eq:prob_profile_1} (Gaussian)
and~\eqref{eq:prob_profile_2} ($n^2\cdot\,$Gaussian) weighted by the transmission factor $T$ from
Fig.~\ref{fig:transmission_factor} (solid lines) compared to the same spectra not weighted by $T$
(dotted lines) for $\alpha=2$.
In both cases the weighted spectrum is heavily concentrated at high frequencies while the lower
components are significantly attenuated with respect to the unweighted spectra, but
the leftmost peaks will persist for substantially longer times than the QNM decay times.
Spectral lines are plotted as Lorentzian profiles with full width
at half-maximum equal to $10^{-2}$ times the decay rate for illustrative purposes.}
\label{fig:grey_spectrum}
\end{figure}

The transmission factor $T=T(\omega)$ is displayed in Fig.~\ref{fig:transmission_factor}.
After weighing the dissipative graviton peaks accordingly, we plot the final spectrum of the dissipation lines
together with the excited QNM lines in Fig.~\ref{fig:grey_spectrum}.
\subsection{The ringing qBH as a leaking cavity}
The power of the low-frequency components of the spectrum is attenuated significantly by the greybody
factor $T(\omega)$, however the energy carried by these gravitons remains in the system.
When a portion of the outgoing flux is scattered backwards and reaches the horizon, we expect
the quantum absorption process will repeat and cause the emission of further gravitons
with energy decreased by a few $\hbar\,\varpi_{\rm BM}$ quanta.
If transmission through the potential barrier is very low and blocks most of the gravitons from escaping,
repetition of this process will cause the entire population of gravitons to drop to the smallest possible
value of energy permitted by the excitation quantum.
These gravitons will no longer be able to drop their energy further and so they will eventually escape
the potential barrier with a very low transmission rate.
\par
This behaviour is analogous to the confinement of soft modes in a cavity with very small leakage.
Since the modes have to escape eventually, being barred from falling inside the horizon in their
entirety, this weak transmission will persist for the large amount of time needed for the disturbance
to dissipate substantially.
We may quantify these statements with a simple estimate of the relevant timescales:
consider first that the time it takes for a light-like signal to propagate from a point at $r = \Rh + \lambda\,\lp$,
where $\lambda$ is a  dimensionless parameter, to the light ring $R_{\rm lr} = 3\,\Rh/2$ is given by integrating
the null line element to obtain
\be
t(R_{\rm lr})
=
R_{\rm lr} - \Rh - \lambda\,\lp+ \Rh\, \ln\!\left( \frac{R_{\rm lr}-\Rh}{\lambda\,\lp} \right)
\ .
\ee
We may take the initial emission point to lie at a Planck length distance from the horizon  ($\lambda=1$)
or several orders of magnitude away as discussed in the context of preparation of asymptotic states
(say, $\lambda\sim 10^{10}$).
The resulting times, for a black hole with $M = 50\, M_\odot$ ($M_\odot$ being the solar mass),
will be respectively $46\,$ms and $34.5\,$ms.
The decay timescale for the $s=l=2$ fundamental mode of a black hole of the same mass will be around $3\,$ms.
It is therefore clear that the signal takes a significantly longer time for a round trip inside the light ring-horizon
cavity than the typical decay time, no matter the choice of $\lambda$.
\par
As a first example, we analyse in details the case with $\alpha=2$ in
Eq.~\eqref{eq:Bek_muk_quantum}, which should be the preferred value according to Ref.~\cite{Bekenstein:1995ju}.
For $M = 50\, M_\odot$, we have $\Rh\, \varpi_{\rm BM} = 0.055$,
and the fundamental QNM with $l=2$ has frequency $\Rh\,\omega_0= 0.747$.
This allows for 13 possible transitions in Eq.~\eqref{P(nE)}.
Each of those will result in an emitted graviton of frequency $\omega_n = \omega_0 - n\,\varpi_{\rm BM}$,
and the latter will propagate as a wavepacket towards the potential barrier which it will cross with
some probability given by the frequency dependent transmission factor $T(\omega_n)$.
We note that the transitions with $n=1$, $n=2$ and $n=13$ respectively have transmission factors
$T(\omega_n)$ around $0.3$, $0.1$ and $10^{-8}$ for $M = 50 \,M_\odot$. 
Even for the case of the highest transmission, this is achieved for about $30\%$
of the emerging flux.
The remainder will therefore reach the horizon at least once more and
its energy will be redistributed to frequencies with even lower transmission rates.
It will take no more than a few tens of round trips (a few $100\,$ms)
for most of the energy carried by dissipative gravitons to be transferred to the lowest
frequency $\omega_{13}$ with the lowest possible transmission $T_{13}\simeq 2\cdot 10^{-9}$.
Assuming then that the energy content $E_j$ of the gravitons at the $j^{\rm th}$ trip is 
\be
E_{j}
=
\left(1-T_{13}\right)
E_{j-1}
\ ,
\ee
this behaviour gives us the approximate expression
\be
E_j
\simeq
e^{-T_{13}\,j}\,E_0
\ .
\ee
The dissipation timescale $t_{\rm d}$ will therefore correspond to $j \sim 1 / T_{13} \sim 5\cdot 10^8$,
leading to
\be
t_{\rm d}
\sim
5\cdot 10^8\, t_{\text{round trip}}
\sim 
5\cdot 10^7\,{\rm s}
\sim
1.5\, {\rm yrs}
\ .
\ee
We conclude that the flux of frequency $\omega_{13}$, though weak, will persist for a timescale several
orders of magnitude longer than the decay time of the QNMs.
\begin{figure}[t]
\centering
\includegraphics[width=\linewidth]{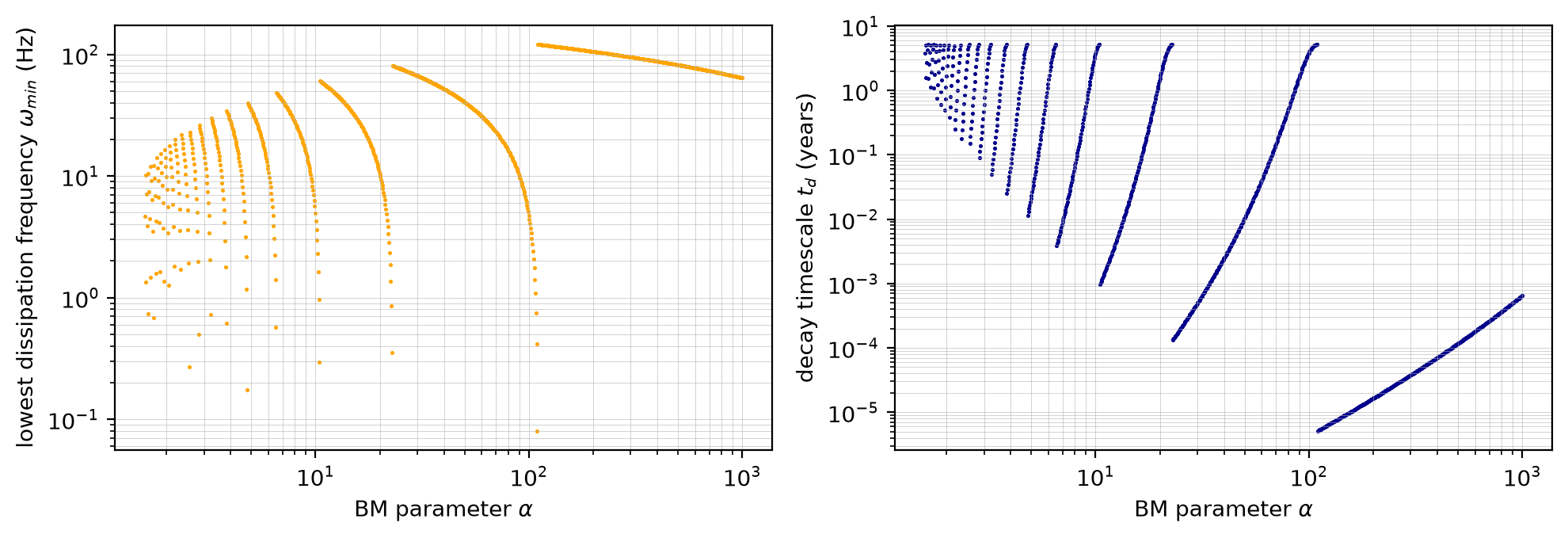}
\caption{Lowest dissipation frequency $\omega_{\rm min}$ corresponding to the fundamental
QNM with $n=0$ for a qBH with $M =  50\, M_\odot$ (left panel) and the corresponding decay timescale $t_{\rm d}$
(right panel) as functions of the quantisation parameter $\alpha$ in Eq.~\eqref{eq:Bek_muk_quantum}.}
\label{fig:decay_timescales}
\end{figure}
\par
In order to assess the validity of this conclusion for more general cases, we repeat the above analysis for a 
whole range of values of the parameter $\alpha>1$ that determines the magnitude of the
quantum~\eqref{eq:Bek_muk_quantum} in different models.
The results for $1<\alpha<1000$ are illustrated in Fig.~\ref{fig:decay_timescales}.
The decay timescale $t_{\rm d}$, in particular, appears to vary within a range of orders of magnitude
that becomes wider as $\alpha$ increases, yet remaining substantially longer than the QNM decay
timescale of $3\,$ms$\,\approx 10^{-10}\,$yrs for values up to $\alpha \sim 10^3$.
The shortest $t_{\rm d}\sim 10^{-5}\,$yrs$\,\approx 300\,$s is reached for $\alpha\gtrsim 100$.  
At the same time, a maximum of $t_{\rm d}\sim 6\,$yrs is reached but not exceeded
for the values of $\alpha$ we explored within a numerical resolution of $\omega > 0.065\,$Hz
(or $\gn\,M\, \omega = 10^{-4}$).
\section{Conclusions}
\label{S:conc}
In this work, we proposed that black holes in the quantum theory maintain the property of absorbing
all infalling objects that reach the horizon, without breaking the classically motivated expectation
that the horizon is in vacuum and represents a causal boundary.
If the black hole mass is quantised, for instance according to Eq.~\eqref{eq:Bek_muk_quantisation},
this necessitates the emission of gravitons carrying the energy increment that is inconsistent with
the transition from one allowed state of the qBH to another.
In this picture, we found that the ringdown spectrum is enriched by the presence of graviton fluxes
with frequencies that correspond to these energy increments, and amplitudes determined by the
relative probabilities of the various excitations that take place. 
\par
These qualitative observations are not significantly influenced by the probability $\mathcal P(n)$
of the qBH transitions in Eq.~\eqref{P(n)}.
In fact, the spectra for the different profiles~\eqref{eq:prob_profile_1}
and~\eqref{eq:prob_profile_2} display similar features in Fig.~\ref{fig:grey_spectrum}
when it comes to the new, low-frequency component.
A detailed calculation of the probability $\mathcal P(n)$, which is not accomplished within this work,
would still be useful for determining the exact spectrum generated from this mechanism.
This specifically requires to identify the relevant interaction vertices that accompany the choice
of asymptotic states and determine the exact dependence of the scattering amplitude on $\delta M \sim n$,
which we here only discussed on general grounds in Section~\ref{SS:scat}.
Furthermore, we only examined even-parity (Zerilli) perturbations of the metric 
because polar and axial perturbations are isospectral~\cite{Chandrasekhar:1975nkd,Chandrasekhar:1975zza},
and extending the analysis to include both types will not influence the expected frequencies
of the dissipative gravitons.
However, their fluxes may still differ, particularly if a physically motivated excitation profile is used,
representing, for example, a perturbed Schwarzschild spacetime following a merger.
\par
From the phenomenological point of view, it is important to remark that estimating the amplitude of
gravitational wave signals corresponding to the predicted flux of soft gravitons requires a precise modelling
of the perturbations that excite the QNM and the subsequent propagation~\cite{Berti:2025hly,RibesMetidieri:2025lxr}.
Such an analysis goes far beyond the scope of the present work.
On the other hand, two solid predictions are that the frequency of such signals is determined by the
qBH quantum $\omega_{\rm BM}$ in Eq.~\eqref{eq:Bek_muk_quantum} and their duration is of the order
of months.
Future experimental data could verify the existence of such a flux or
allow us to place bounds on $\omega_{\rm BM}$, hence on the mass spectrum of qBH.
In particular, the frequency of soft gravitons should fall within the sensitivity band of the third-generation
detectors~\cite{ET:2025xjr}.
The predicted signal could last from a few hundred seconds up to approximately $6\,$yrs,
with a preferred duration of about $1.5\,$yrs (obtained for $\alpha=2$) according to the original argument
of Ref.~\cite{Bekenstein:1995ju}.
For long durations, we can consider the signal to be nearly monochromatic.
In this case, the appropriate techniques to analyse signals over such long timescales, on the order of a year,
are those developed for continuous wave search, ranging from semi-coherent methods~\cite{Mirasola:2024kll},
in which the data are divided into segments that are analysed coherently and then combined incoherently,
to fully coherent integration over the entire emission.
The two extremes of this range represent a trade-off between sensitivity and computational cost,
but in both cases the signal-to-noise ratio is averaged over the entire duration of the signal.
A simulation study that incorporates these signals into the simulated noise of a third-generation
detector is currently in preparation to evaluate the detectability of the effect as a function of
$\omega_{\rm BM}$ and the emitted flux.
\par
Finally, we remark that a more careful examination is also warranted for the regularising cutoffs required
to amend the logarithmic divergence of $\expec{\hat N}$ in Eq.~\eqref{divN}, through which the
correspondence with the Bekenstein-Mukhanov was posited in Eq.~\eqref{reg_ratio_to_Bek_Muk}.
While some form of regularisation should be in place for our qBH is constructed to resolve
known pathologies of the classical solution, it remains to clarify whether there exists
a regularisation scheme that is fully consistent with the Bekenstein-Mukhanov
law~\eqref{eq:Bek_muk_quantisation}.
\subsection*{Acknowledgements}
R.C.~is partially supported by the INFN grant FLAG and carried out this work in the framework
of the activities of the National Group of Mathematical Physics (GNFM, INdAM) and the COST
Action~CA23115 (RQI).
\subsection*{Data availabiliy}
The code employed to generate the results of this study is available at
\url{https://github.com/KonTopal/Quantum_ringdown/tree/master}.
\end{document}